\documentclass[12pt,preprint]{aastex}

\usepackage{epsfig}
\usepackage{amsmath,amsfonts,amssymb}
\usepackage{subfigure}
\usepackage{comment}
                                                                                                    
\newcommand{\el}{\ensuremath{e^-}}

\renewcommand{\fh}{\ensuremath{^{h}}}
\renewcommand{\fm}{\ensuremath{^{m}}}
\renewcommand{\fs}{\ensuremath{^{s}}}

\newcommand{\ctbd}[1]{}

\slugcomment{DRAFT}

\shorttitle{Millimagnitude photometry with HAT for Extra-solar Planet
Detection}
\shortauthors{Bakos et al.}

\begin{document}

\title{Wide-field millimagnitude photometry with HAT: \\
A Tool for Extra-solar Planet Detection}

\author{G.~Bakos\altaffilmark{1,2,3}
R.~W.~Noyes\altaffilmark{1}, 
G.~Kov\'acs\altaffilmark{2},   
K.~Z.~Stanek\altaffilmark{1},  
D.~D.~Sasselov\altaffilmark{1} and
Istv\'an Domsa\altaffilmark{2}
}
\email{gbakos,rnoyes,kstanek,dsasselov@cfa.harvard.edu; 
kovacs,domsa@konkoly.hu}

\altaffiltext{1}{Harvard-Smithsonian Center for Astrophysics,
60 Garden street, Cambridge, MA02138}
\altaffiltext{2}{Konkoly Observatory, Budapest, H-1525, P.O. Box 67}
\altaffiltext{3}{Predoctoral Fellow, Smithsonian Astrophysical Observatory}

\begin{abstract}
We discuss the system requirements for obtaining millimagnitude
photometric precision over a wide field using small aperture short
focal length telescope systems, such as are being developed by a number
of research groups to search for transiting extra-solar planets.  We
describe the Hungarian-made Automated Telescope (HAT) system which
attempts to meet these requirements.  The attainable precision of HAT
has been significantly improved by a technique in which the telescope
is made to execute small pointing steps during each exposure so as to
broaden the effective point spread function of the system to a value
more compatible with the pixel size of our CCD detector.  Experiments
during a preliminary survey (Spring 2003) of two star fields with the
HAT-5 instrument allowed us to optimize the HAT photometric precision
using this method of PSF broadening; in this way have been able to
achieve a precision as good as 2 millimagnitudes on brighter stars. We
briefly describe development of a network of HAT telescopes (HATnet)
spaced in longitude.
\end{abstract}


\section{Introduction}
\label{sec:intro}

The discovery of the transiting planet \mbox{HD 209458b}
\citep{DC00,Henry00} was a landmark in exoplanet research. It
established that at least some substellar companions to stars found in
close orbits ($\sim$ 0.05 AU) by radial velocity techniques really are
gas giants planets, with mass and radius comparable to theoretical
expectations for a ``hot Jupiter'', that is, with radius of order 35\%
larger than expected for a Jupiter-mass gas giant planet orbiting at
several AU from the star. Because \mbox{HD 209458} is a relatively
close and bright star (I band magnitude 7.0), detailed follow-up
observations were possible, and those from {\em HST\,} \citep{TB01}
yielded the first measurements of its atmospheric sodium composition,
showing it to be broadly consonant with theoretical expectations
\citep{DC02}; the existence of a hydrogen exosphere \citep{VM03}; as
well as detailed stellar-planet system parameters \citep{Queloz00}.
Thus the star's brightness has allowed us to explore the physics of an
extrasolar planet atmosphere years ahead of expectations. In
recognition of this, we have embarked on a survey for additional
examples of planets transiting nearby bright stars, using a small
Hungarian-made Automated Telescope (HAT) system.

Many projects searching for transiting giant planets have sprung up
since the discovery of \mbox{HD 209458b}, and can be divided in two
main types: deep and narrow surveys use medium to large size telescopes
with a narrow field-of-view (FOV) to record many faint sources, while
wide and shallow surveys use small telescopes with a wide FOV to record
a great number of bright nearby stars \citep[see][for a comprehensive
list of such projects]{KH03}. The charm of small telescopes is that
instrumentation is affordable, even allowing building dedicated
instruments for the project and thus, assuming robust automation,
having practically unlimited telescope time. Furthermore, if a
transiting candidate is found, the brightness of the parent star
enables prompt follow-up observations (both photometric and
spectroscopic) with 1m-class telescopes to determine whether the
candidate is worth further investigation with large and unique
facilities (10m class telescopes and {\em HST\,}), or whether it is a
false positive mimicking a transit for other reasons. However, the
results so far have been disappointing, since (as of this writing) no
detections have yet been announced using these wide-field surveys.
While this has given rise to some skepticism about whether they will
ever bear fruit, this skepticism is unwarranted; it is simply the case
that earlier projections of the detection rate \citep[e.g.][]{KH03}
were overly optimistic \--- perhaps by as much as a factor of 10
\citep{TB03}. In fact, the only planet discovered by the transit method
(OGLE-TR-56b), and not by radial velocity searches, belongs to the
``deep and narrow'' surveys, as it was found with the 1.3m OGLE
telescope at $I\approx 16$ \citep{Udalski02,Konacki03,Torres03}.


A useful discussion  of expected detection and false alarm rates for
transiting Jovian planets has been given in \citet{TB03}. These rates
can be calculated by combining two terms. The first is an empirical
estimate on the probability that a star with magnitude $m_0$ (and color
index $C_0$) harbors a transiting Jovian planet with period P,
fractional transit duration q and photometric depth $\delta$. This
distribution function also depends on some secondary parameters such as
galactic latitude $b$ and longitude $l$, which relate to the
distribution of stellar types (hence radii), as well as field crowding
and thus false alarm rates.
The second term comes from the capabilities of the survey: photometric
precision $\sigma$ as a function of $m_0$, number of data-points
accumulated for a star and their distribution in time, and angular
resolution of the imaging.
For completeness, there is also a third term, which is the data
reduction method used for filtering the transit candidate light-curves
from the flood of data. 

The combination of these terms yields a joint probability distribution,
which tells what the chance is that a single star ($m_0$, $C_0$, $l$,
$b$) has a bona-fide transit (with characteristics P, q, $\delta$) {\em
and} gets selected when the specific transit-search algorithm is run on
the accumulated data (with given number of data points and time
distribution) collected by the survey that is characterized by its
photometric capabilities. Naturally, the above joint probability has to
be integrated over the number of stars observed by the survey. In this
paper we concentrate on the ``observational'' term of the probability
distribution, i.e.~on how to optimize the characteristics of our
ongoing HAT survey in order to maximize the number of expected
detections, with special focus on the optimizing photometric precision
through a PSF-broadening (PB) technique.

Throughout this paper, by (photometric) ``precision'' we will refer to
the repeatability of the measurements for the same star, i.e.~the rms
of the JD versus magnitude time-series (light-curve).  The light-curves
are established by relative photometry, essentially using differential
magnitudes based on most of the bright stars in the field. Because of
the nature of the search for photometric variations that are not known
a priori, a great number of such stars are used as comparison stars
through an iterative and robust procedure of rejecting outliers, and
finding the frame to frame magnitude offsets. For clarity, we
distinguish between the repeatability as described above, and the
accuracy of the relative magnitude measurements within a single frame.
The latter is characterized by the difference between measured
magnitude values for the same virtual star if placed at different XY
positions on a single frame. Finally, accuracy of the standard
photometry describes the relation of the instrumental magnitude system,
which is characteristic to the project, to the standard photometric
system, such as that of \citet{Landolt92}.

Given the expected depth of only $\sim 1\%$ of hot Jupiter transits,
photometric precision of time-series measurements plays a critical role
in detection capabilities \citep[][Fig.~3]{TB03}. However, achieving
adequate ($\le 1\%$) precision of the light-curves over a wide FOV
projected on a front-illuminated CCD, and on data coming from an
unattended automated instrument, poses serious technical challenges,
which might be partly the reason for the negative results so far of
wide-field surveys.  These challenges are discussed in \S
\ref{sec:photchal}.

In order to make HAT suitable for extra-solar giant planet transit
detection, and improve our precision, several hardware modifications
were performed on the prototype HAT-1 system \citep{GB02}. Section \S
\ref{sec:hw} gives an overview on these, and \S \ref{sec:pb}, describes
our PB-technique that has improved our precision still further. The
upgraded, new-generation HAT, called HAT-5, was started up early 2003
at the Smithsonian Astrophsical Observatory's Fred L.~Whipple
Observatory (FLWO), Arizona, and has been running since. Sections \S
\ref{sec:obs} and \S \ref{sec:red} summarize our observations taken in
the Spring season of 2003 with HAT-5, and data reduction steps relevant
to achieving the photometric precision.  We compare our photometry
taken with standard ``tracking'' mode and the PB-technique in \S
\ref{sec:photprec}, and explain why PB yields better performance. The
resulting short and long-term photometric precision, as well as
systematic variation in the light-curves are discussed in the same
section. In \S \ref{sec:res} we describe our experience with the BLS
transit search method based upon the 91-night dataset of the above
Spring season, and also show sample light-curves. Finally, in \S
\ref{sec:sum}, we summarize the precision obtained with HAT-5, and
touch upon future prospects of the HATNet - a multi-element network
consisting of identical instruments, which has been operational since
November, 2003.  Further information can be found at the HAT
homepage\footnote{http://www-cfa.harvard.edu/\~{ }gbakos/HAT/}.

\section{Photometric precision with a wide-field, short focus instrument}
\label{sec:photchal}


As stated earlier, photometric precision of time-series measurements is
a key factor in detection capabilities. Although the events searched
for are periodic, the loss in precision (characterized by $\sigma$, the
rms observed magnitude variation of a constant star) is difficult to
overcome by increasing the number of data points, $N_{obs}$: the
detection chance is roughly proportional to $\sqrt{N_{obs}}/\sigma$.
Furthermore, the number of data points per star that can be accumulated
(in a year) is obviously limited by the seasonal visibility of fields.
Improving the precision per data point rather than increasing $N_{obs}$
 facilitates detecting a potential transit early in an observing
season, so as to allow follow-up observations that same season.
Naturally, photometry at the level of 1\% or better can contribute
strongly to general stellar variability studies in addition to
exoplanet transit searches, for it enables study of otherwise
undetectably low-amplitude variables.

Achieving precision better than 1\% over an extended 
FOV poses serious technical challenges. The following complications
that arise in astrometry and photometry for a typical, low-budget
planet search project are marked with two-character symbols for future
reference (see Table~\ref{tab:issues} for a summary).

{\it Wide Field Issues (W):} The requirement of a wide FOV in order to
observe a large number of bright sources translates into short focal
length for moderate sized CCD detectors.  This plus the need for
reasonable aperture to maximize the incoming flux, results in rather
fast focal ratio instruments (for example f/1.8 lenses in the HAT
project). As a result, the geometry of the focal plane is significantly
distorted, so that astrometry is more complicated than with
``conventional'' telescopes (W1). In addition the stellar profiles are
far from perfect, and change significantly over the wide FOV (W2). This
gets worse with even slight defocusing, and the fast focal ratio
increases the difficulty of achieving satisfactory focus. (W3). Even
matching frames of the same instrument can be problematic due to
differential refraction (W4), e.g.~a corner of the field is
``lifted-up'' at high zenith angles. Finally, differential extinction
(W5) across the field can be considerable at high zenith angles.

{\it Flat-fielding Issues (F):}  
Commercially available telephoto lenses exhibit strong vignetting (both
optical and geometrical), typically having only 60-80\% incoming
intensity in the corners compared to that of the center of the field.
While this is in principle corrected by flatfielding, residual
large-scale flatfielding errors are proportional to the amount of the
correction and hence can be significant (F1). The many lens elements
(the HAT lenses consist of 12 lens elements in 10 groups) cause
reflected stray-light patterns (F2) to appear on the images, and
further decrease flatfielding precision. The sky background is variable
(F3) over a typically few degree scale, both spatially and in time
\citep{Chromey96}, so that skyflats do not truly represent the
transmission function of the system on large-scale, and median
combining is problematic, so that small-scale errors are more enhanced
than at large telescopes (F4).  Due to the short focus, and
depth-of-field (close objects are not completely out of focus), it is
virtually impossible to find any setup for domeflat exposures.

{\it Sky Noise and Undersampling Issues (S \& U):} 
With typical CCD pixel sizes of $\sim 14\mu$ and focal lengths of
$\sim200\,mm$, the pixels correspond to large area on the sky
(e.g.~$14\arcsec\times14\arcsec$), and hence sky-background is one of
the major contributors to noise (S1). For wide field short focal length
lenses such as we employ, the optical point spread function (PSF) has a
full width at half maximum (FWHM) of order 20\arcsec.  Hence the
stellar profiles are undersampled on the CCD chip. This involves
several further complications, the most prominent being the increased
effect of residual small-scale flatfield errors in the photometry (U1).
Most methods for finding the centers of sources start to break down
below $\sim$2\,pix FWHM (U2). Undersampled profiles cause problems for
both PSF-fitting (U3), and flux-conserving image interpolation (U4) to
a reference frame. Finally, the maximum star brightness before pixel
saturation occurs at fixed exposure time is less if the PSF is narrow
and undersampled (U5).  This limits photometric precision due to
additive (per frame) noise-terms, such as readout-noise.

{\it Merging Issues (M):} 
Merging of stellar sources is increased not only by the fast focal
ratio and limited resolving power of the optics, but also by the
undersampled profiles (M1). While at first approximation, a constant,
time-independent merging of sources on the frames would not harm the
photometric precision of a time-series measurement, any intrinsic
variation exhibited by one of the stars (e.g.~a shallow transit) is
suppressed by the flux contribution of nearby sources. However, even a
constant merging affects the precision through centroid and
sky-background determination errors.  Furthermore, even with perfect
telescope pointing, if the PSF and hence merging of sources is variable
in time, then the the relative brightness of stars varies from frame to
frame (M2).

{\it Other CCD Issues (C):} 
If only semi-professional (but affordable), front illuminated CCDs are
used, this poses further difficulties.  For example, the HAT project
uses Apogee AP10 2K$\times$2K, front illuminated camera with 14$\mu$
pixels. The gate and channel structure in front of the pixels causes an
unknown intra-pixel (C1) and inter-pixel (C2) variability
\citep{Buf91}. The dark-current with such, typically Peltier-cooled
systems is not negligible (C3).

%
Depending on the goals of the photometry, and the stability of the
instrument, such as pointing precision and focusing stability,
different subsets of the above effects are important. In the following
we distinguish between ``major'' (few tens of pixels), ``minor'' (a few
pixels) and sub-pixel (better than few tenths of a pixel) pointing
accuracy, and nightly or long-term photometric precision. Some of the
effects, such as S1, M1 and U2-5, are simply present in general,
irrespectively of the above categories.

If one performs {\em absolute astrometry} of the fields, i.e.~cross
identifies stars with a catalogue, then W1, W4, U2, U3 and M1-2 have to
be dealt with. 
{\em Relative astrometry} between frames from the same instrument will
generally depend on the same effects, except for dependence on
distortion of the optics (W1), if the pointing errors are ``minor'' or
better.

{\em Relative photometry} is the typical application needed by
transit-search projects to generate time-series of stars.  In an ideal
setup, if the fields were observed with sub-pixel pointing precision
and constant PSF width at the same night (and reduced with the same
flatfield), the photometric precision (but not the accuracy) would be
almost indifferent to all the problematic effects except those
generally present for photometry (W5, S1, M1, U2-5).  However, because
W4 (differential refraction) causes displacement of the centroids, even
at perfect tracking and polar alignment, sub-pixel positioning is
possible only in a limited part of the field, and over a limited hour
angle range. Thus relative photometry with even excellent pointing
(e.g.~through auto-guiding) is still slightly affected by F4, U1, C1-2,
and with increasing pointing errors their contribution is amplified. If
pointing accuracy is even worse than ``minor'' (i.e.~a few tens of
pixels), then virtually all the remaining effects have to be dealt
with. If focusing stability is poor, which is usually relevant to
long-term precision, then variable merging (M2) is also a significant
contributor to noise in crowded fields.

Finally, {\em absolute photometry} and calibration of fields to a
standard system suffers from almost all the above problems. For example
large-scale flatfielding errors (F1), field-position dependent profiles
(W2), and stray-light (F2) can cause up to $\sim 5\%$ absolute
calibration errors.

\footnotesize
\begin{table}[!h]
\begin{tabular}{ll}
Issue & Short explanation\\\hline\hline
\multicolumn{2}{l}{Wide Field}\\\hline
W1	&	Distorted geometry in astrometry\\
W2	&	Spatially variable stellar profiles\\
W3	&	Sensitivity of profiles to focusing\\
W4	&	Differential refraction\\
W5	&	Differential extinction\\
\multicolumn{2}{l}{Flat-fielding}\\\hline
F1	&	Strong vignetting, uncertainty of large corrections\\
F2	&	Stray-light patterns\\
F3	&	Variable sky background\\
F4	&	Increased small-scale flatfield errors\\
\multicolumn{2}{l}{Sky Noise and Undersampling}\\\hline
S1	&	High sky background per pixel\\
U1	&	Increased effect of residual flatfield errors\\
U2	&	Imprecise centroid fitting\\
U3	&	Breakdown of psf fitting\\
U4	&	Problematic image interpolation\\
U5	&	Limited flux for brightest unsaturated stars\\
\multicolumn{2}{l}{Merging}\\\hline
M1	&	Increased merging\\
M2	&	Time-dependent merging due to change in profiles\\
\end{tabular}
\caption{Issues in wide-field astrometry and photometry\label{tab:issues}}
\end{table}
\normalsize

\clearpage
\section{History and current hardware}
\label{sec:hw}

\begin{figure}[!h]
\begin{center}
\epsscale{0.8}
\plotone{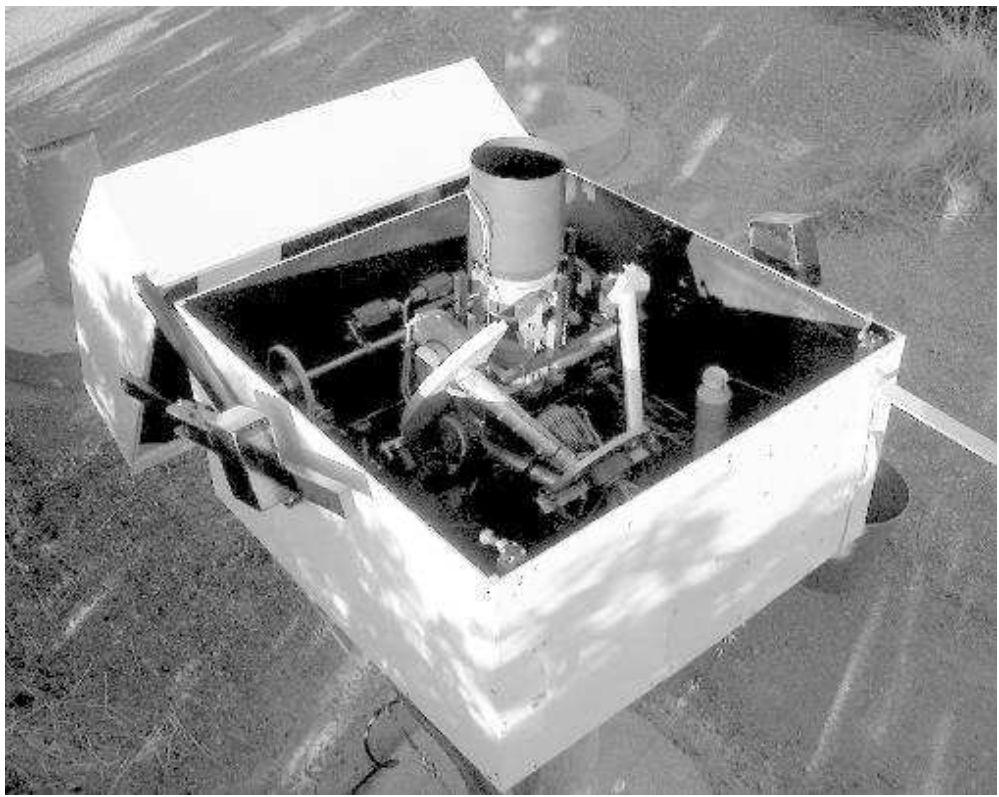}
\caption{The new generation HAT-5 installed to FLWO: Canon 200\,mm f/1.8
lens and Apogee 2K$\times$2K CCD attached to a friction driven
horseshoe mount, protected by a clamshell dome.
\label{fig:hat5}
}
\end{center}
\end{figure}

The development of HAT was initiated by Bohdan Paczy\'nski in 1999 with
the original purpose being all-sky variability monitoring. HAT is a
small automated observatory, incorporating a robotic horseshoe mount, a
clamshell dome, a large-format CCD, a telephoto lens, and auxiliary
devices, all controlled by a single PC running RealTime Linux. Except
for the CCD and lens, all the components, including the software
environment, were designed, developed and manufactured by our team in
Hungary. The design of the horseshoe mount was based on the ASAS
instrument \citep{GP97}. The prototype instrument, HAT-1, was
operational for more than a year (2001/2002) at Steward Observatory,
Kitt Peak, AZ. The typical photometric precision we reached with an
Apogee AP10 front-illuminated, 2K$\times$2K CCD, 6cm diam f/2.8 Nikon
lens and I-band filter flattened out at 1\% for the brightest stars
($I\sim6.5$). Further details of the instrument and the startup period
can be found in \citet{GB02}.  HAT-1 was decommissioned in the fall of
2002 to become part of an upgraded system, the new generation HAT (the
prototype called HAT-5, see Fig.~\ref{fig:hat5}). Here we restrict
ourselves to brief summary of the modifications we performed in order
to improve the precision, motivated by planet transit searches.

We were fortunate to acquire by long-term loan the hardware of the
de-comissioned ROTSE-I project \citep{ROTSEI,Akerlof00}. The CCDs were
identical to the one we already used, but the Canon 200\,mm f/1.8L
lenses have a four times larger entrance pupil, and are of much better
optical quality than the previously used lens. Many hardware
modifications were made, in order to ameliorate difficulties of precise
relative photometry over the wide field of view, as described in \S
\ref{sec:photchal}.

The HAT horseshoe mount, an open-loop control system, is
friction-driven by stepper motors, without encoders to record the
exact position. Our pointing accuracy was improved by re-designing the
horseshoe and the declination drive, thus achieving better friction and
balance of the telescope. Use of precision rollers for the RA-axis
decreased our periodic errors in tracking. We do not use auto-guiding
for several reasons: i) it complicates the current, relatively
inexpensive hardware setup, ii) our tracking errors in 5 minute
exposures are negligible, iii) due to differential refraction, stars in
the corners would drift away on a sub-pixel scale even with
auto-guiding. However, we are able to perform astrometry immediately
after each exposure, and correct the position of the telescope before
the next exposure, thus achieving few-pixel pointing precision.

The fast focal ratio lens also involved several modifications. We had
to develop a computer-controlled focusing system, driven by a 2-phase
stepper motor (\S \ref{sec:photchal} W3). The different lens, and
longer effective ``telescope'' length resulted in re-designing the
entire dome, and lens-supporting mechanism.

The effects of differential extinction (\S \ref{sec:photchal} W5) can
be minimized by use of multiple filters, and inclusion of color-terms
in the fits of magnitude differences between individual frames.
Multiple filters are also essential for quick elimination of some false
transiting planet signatures, and for proper standard calibration of
the photometry. Because one of the purposes of the HAT instrument is to
provide useful data for variability monitoring as well, we decided to
use standard Cousins I filter as primary band, complemented by Johnson
V (both made following the prescription of \citealt{Bessel90}). The
14$\mu$ pixel size, fast focal ratio and short backfocus of the lens
(42\,mm) pose stringent requirements on the filter exchanger. We
designed our own exchanger which is capable of few micron
re-positioning accuracy.

\section{PSF broadening}
\label{sec:pb}

The hardware modifications outlined in \S \ref{sec:hw}, and the
custom-built software (see \S \ref{sec:red} below) took care of some of
complications that arise from wide-field photometry, but not of effects
that are related to the undersampled profiles (see \S
\ref{sec:photchal}: F4, U1-5, C1-2). The stellar profiles of HAT-5 are
$\sim$1.5\,pix, at pixel scale of 14\arcsec, showing slight variation
over the 8.2\arcdeg$\times$8.2\arcdeg FOV. Our suspicion was that
broadening the instrumental PSF, so as to achieve better sampling,
could improve the photometric precision, which previously was never
less than $\sim$1\%.

Because we did not have the freedom to change the CCDs to a type with
finer pixel-resolution, achieving better sampling required widening the
stellar profiles. Our first attempt to do this by generating
artificially bad seeing in front of the lens with heating elements
failed.  Defocusing of the lens enough to broaden the profiles
sufficiently did not work, because it introduces strong, spatially
dependent distortion of the profiles.  Hence we decided to broaden the
PSF by stepping the telescope pointing during the exposure.

\begin{figure}[!h]
\begin{center}
\epsscale{0.45}
\plotone{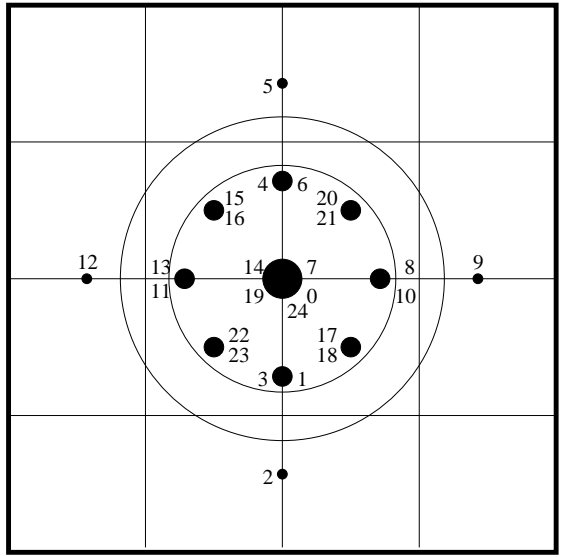}
\caption{
The PSF-broadening pattern the telescope executes during an exposure.
The amplitude of tiny movements is 10\arcsec, which corresponds to 10
microsteps in RA (on the Celestial Equator), and 2 microsteps in Dec.
The pattern starts from the center to the South, continues to the N, W,
E, and finally ends in the diagonal directions, as indicated by
consecutive numbers. The size of dots is proportional to the dwell
times spent at that grid-point. The solid-line grid shows the pixel size
(14\arcsec), the inner concentric circle the FWHM of a typical
intrinsic PSF (1.7\,pix), and the outer circle the FWHM of the
broadened PSF (2.3\,pix).
\label{fig:pbpat}
}
\end{center}
\end{figure}

The resolution of the HAT stepper motors is 1\arcsec\ in RA (1/15\,pix)
and 5\arcsec\ in Dec (1/3\,pix), respectively, and any sequence of
stepping commands can be generated and superposed on the sidereal-rate
tracking from software, using the RealTime Linux drivers. We
reprogrammed the high-level image acquisition program so that during
the exposure the telescope steps around the central position on a
prescribed pattern (Fig.~\ref{fig:pbpat}).

We experimented with $\sim$25 different broadening-patterns, that
is, patterns of pointing offsets from the central pointing position,
together with dwell times at each pointing offset. Small movements of
the telescope were found not to be completely deterministic, because of
missing microsteps and the elasticity of the sprockets, so we perform
the stepping pattern a few times (generally 3) during the exposure to
avoid asymmetries in the profiles. We also modeled the expected profile
by superposing the intrinsic 1.7\,pix wide Gaussian profiles on the
offset pointing grid with weights corresponding to the time-intervals
spent at the grid-points. When composing a wider profile by this
superposition, too large-amplitude grid-steps cause a flat-top profile
with humps, i.e.~the resolution of the grid has to be smaller than the
intrinsic FWHM. We require that the superposed profile also be a
Gaussian to a very good approximation; only a few of the patterns
fulfill this requirement.
The best pattern we achieved consisted of stepping during the exposure
to positions 10\arcsec\ and 20\arcsec\ in a N,E,S,W direction from the
center position, and positions 10\arcsec\ from the center in a NE, SE,
SW, and NW direction (see Fig.~\ref{fig:pbpat}). The resulting PSF is
2.3\,pix wide, broadened from the 1.7\,pix intrinsic value with no
stepping.

The great advantage of this PSF-broadening approach is that to first
order all the stars in the FOV are broadened in the same manner. To
second order, because the telescope is moved in the RA-Dec system,
stars experience slightly different true angular movements in RA as a
function of declination. The ratio of these movements on the top and
bottom of the 8.2\arcdeg\ field is close to unity, being exactly 1.0
when the field center is at $\delta=0\arcdeg$, and reaches 1.18 at
$\delta\approx50\arcdeg$, although the ratio of resulting profile
widths is less. This is approximately our tolerance 
(defined by being equal to the nightly variation
of the PSF in the center of the field we experience due to other
reasons), and hence sets a limit on observational declination. 

PB not only broadens the profiles, but also smooths out their
spatially-dependent distortion, producing profiles that are more
homogeneous over the field. Also, while the intrinsic profiles are
subject to tracking errors and other impacts (gusts, etc.), the PB ones
are more stable in time.
This technique yielded far better precision for bright stars in
moderately crowded fields with simple aperture photometry than the
``tracking'' frames, as described in \S \ref{sec:photprec}.  We
emphasize that it is not the only way of improving precision for
small-telescope projects, but rather a specific practical solution
given our restrictions (CCDs, lenses, budget).

\begin{figure}[!h]
\begin{center}
\epsscale{1.0}
\plotone{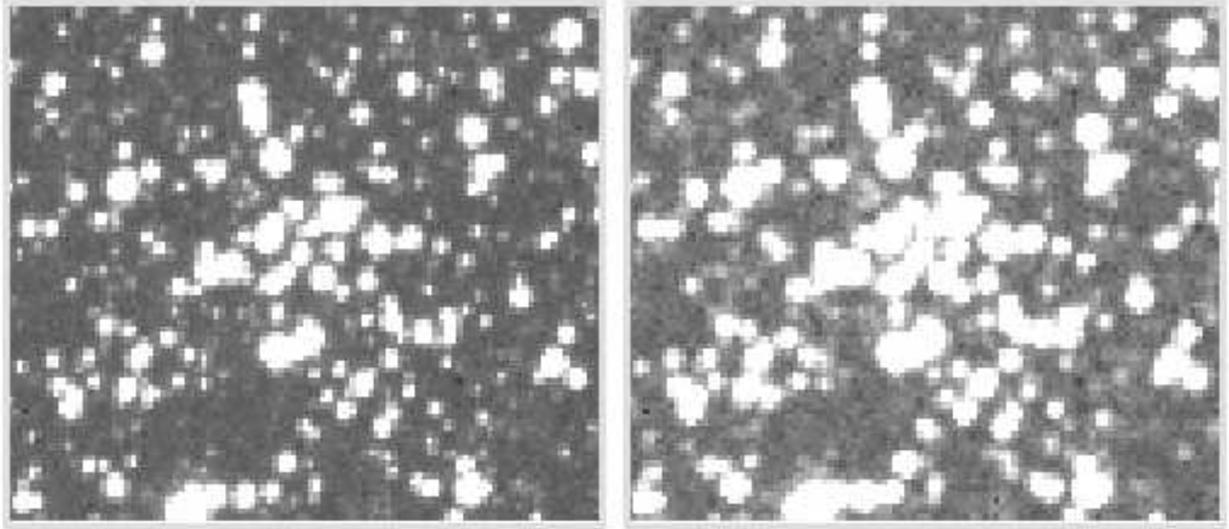}
\caption{
A $25\arcmin\times25\arcmin$ region (approx 1/300th of a frame) showing
the open cluster M34 taken by HAT-6 with tracking mode (left, 1.2\,pix
FWHM) and PSF broadening mode (right, 2.0\,pix FWHM). Blurring
is to the eye only slight, but makes a significant difference in
photometry. 
\label{fig:pbimg}
}
\end{center}
\end{figure}

\section{Observations}
\label{sec:obs}

HAT-5 started up observations at FLWO in February, 2003. The first
month was spent with various experiments, such as fine-tuning the PB
patterns. Routine observations started in March, 2003, and the
telescope observed on 91 nights until the monsoon arrived (10 July).
Although HAT-5 was started up again after the monsoon (in September),
now as part of the HAT Network, along with two other telescopes at
FLWO, the primary subject of this paper is the data that have been
reduced and analyzed from the Spring season of HAT-5.

The typical scenario at each observing session was the following: in
the early evening, after cooling down the CCD, bias frames were taken.
If the sky was clear, as seen from a live webcamera and satellite
images, the telescope was enabled to open up (remotely, over the
Internet, typically from the CfA). This is the only point that needed
manual interaction with the telescope system; otherwise the entire
observing session and the scheduling of tasks was done automatically.
The skyflat program started at sunset, and frames were taken at the
closest flatfield region to the optimal point on the sky with the
smallest gradients \citep{Chromey96}. The main program was launched by
the automatic ``virtual observer'' program after nautical twilight. We
observed two selected regions; one sparse field in the beginning
of the night (in the constellation Sextans, centered on $\alpha$ =
10\fh00\fm00\fs, $\delta$ = 00\arcdeg00\arcmin00\arcsec, and labeled
G416), and a moderately crowded field in the second half of the night
(in the constellation Hercules, centered on $\alpha$ = 17\fh36\fm00\fs,
$\delta$ = 37\arcdeg30\arcmin00\arcsec, and labeled G195). Both fields
were observed with 300s exposures, primarily in I-band, and with the
PB-pattern that yielded $\sim 2.5$\,pix FWHM on average. Unfortunately,
due to a slow drift of the instrument adjustments, as well as seasonal
thermal expansion, the PSF increased from 2.3\,pix (early Spring) to
more than 3\,pix (June). Occasionally tracking-frames were taken, as
well as V-band images. Depending on the length of the night, approx.~80
frames were accumulated per session, with equal number of G416 and G195
frames in the beginning, and almost entirely G195 frames at the end of
the Spring season. The maximal zenith angle of the fields was
60\arcdeg. Immediate post-exposure coordinate corrections (see \S
\ref{sec:hw}) were not applied; their potential importance was only
realized later, when reducing this dataset.  During the entire season,
we collected 1300 frames for field G416 and 3300 for G195. The morning
skyflats were taken in a similar manner to the evening. We closed the
session by taking long dark frames and bias exposures. All data were
archived to tapes immediately after the session.
For comparison of the tracking and PB-patterns, some nights were
devoted to experiments, in which frames of the same field were taken in
alternating mode. This way we could directly compare the precision of
the two methods by looking at the rms of light-curves from tracking and
PB light-curves.

\section{Data reduction}
\label{sec:red}

Image calibration was performed in a standard way, using GNU/Linux
shell scripts to control IRAF\footnote{IRAF is distributed by the
National Optical Astronomy Observatories, which are operated by the
Association of Universities for Research in Astronomy, Inc., under
cooperative agreement with the National Science Foundation.} and
self-developed tasks. Each night is dealt with as a separate unit with
its own calibration frames, unless some of the calibration observing
programs failed (bad weather, no useful skyflat regions, etc.). In
these cases, calibration from neighboring nights was used. Standard
overscan, bias, dark and flatfield corrections were applied. (The
Thomson chip in the AP10 CCD has no true overscan; instead we are using
an electronic overscan, which is the read-out of several columns before
the image columns.) Inferior quality skyflat frames are rejected using
various fits to statistical parameters of all the flats taken during
the session (standard deviation vs.~mean value, proximity to desired
8000\,ADU mean level, etc.). This way the presence of clouds in the
skyflats can also be detected, in case they were not checked or
realized on the web camera. The images are also subject to quality
filtering before beginning the astrometry and photometry, using mean
level, standard deviation of the background, profile widths, etc.

\paragraph{Astrometry} 
We used fixed-center aperture photometry to derive instrumental
magnitudes. The central X,Y positions of the sources on any frame are
those of an astrometry reference frame (AR) transformed to the
individual frame. The AR is chosen to be a tracking frame, where
profiles are sharp, and merging of the sources is less severe than on
PB images. (Note, therefore, that the full implementation of the PB
technique requires the acquisition of both tracking and PB-frames.) A
grid of $\sim$2000 stars is used on the AR and individual frames to
determine the geometric transformation between the frames.  We start by
triangle matching the frame to the AR grid, and after a linear
transformation is established, we increase the rank of the fit, and
tune the sigma clipping, until we reach a satisfactory 4th order fit
with rms of residuals smaller than 0.1\,pix.  This iterative procedure
is designed to work in a more general application of matching frames
with a star catalogue, and is more complicated than necessary for
simply matching frames to each other, but was used for convenience.
General triangle matching is extremely CPU-intensive ($\propto
N^{9/2}$, where N is the number of stars), because a great many
($\approx N^3/6$) triangles are generated and have to be paired
\citep{Groth86}. Furthermore, in this case the larger triangles can be
significantly distorted with such a fast focal ratio imaging. So
instead, we generate a mesh of local triangles with an optimized
algorithm, using Delaunay-triangulation \citep[e.g.][]{Del96}, and use
these local triangles (only N-2 per frame) for finding the initial
transformation. The computation with Delaunay-triangulation scales only
as $N^p$, where $2 < p < 3$. While Delaunay-triangulation is a more
special case than general triangle-matching algorithms (because it
requires similar surface density of the two datasets to be matched, so
that the local triangles generated have a common subset), in our case
matching the frames to each other always works, except for frames with
defects. The same holds for matching frames with the Guide Star
Catalogue \citep{GSC}.

\paragraph{Photometry}
Fixed center aperture photometry was performed on all the PB frames
using the centers we got from transforming a ``modified'' AR coordinate
list to each frame. This modified AR was derived by first selecting one
PB frame as the photometric reference (PR) from a dark, photometric
night. Then the transformed coordinates of the brightest, non-saturated
stars that were missing from the AR (because they saturated on the
tracked frames) were appended to it.

The instrumental magnitudes vary between the frames for a number of
reasons. First, the extinction changes from night to night, with zenith
angle, or with variable photometric conditions. Second, the profile
widths also vary, changing not only the accumulated flux in the fixed
aperture but also the merging with neighboring sources (\S
\ref{sec:photchal}: M2). Depending on accuracy of field-positioning
(``minor'' or ``major''; see \S \ref{sec:photprec}), the different
placement of the star also causes magnitude offsets (``minor'': F2, F4,
U1-2, C1-2; ``major'': these plus W2 and F1). To correct for these and
possibly other factors, we fitted the magnitude differences from the
reference PR frame of the brightest thousand stars as a function of X,Y
coordinates in an iterative way with 3 sigma outlier rejection, so
as to eliminate variable sources from the magnitude fits. 
Color-dependence was not incorporated in the fits. The residual of the
fits (to fourth order in X and Y) can be as low as 3 millimag, and was
typically better than 6\,mmag. For non-photometric nights the fit rms was
significantly worse than 1\%; these large rms values may be used as a
criterion to reject such nights.  All parameters of the fit
were recorded, and later used for quality filtering the data-points.

Based upon the rms of the magnitude fits, approximately 20 of the best
frames from the same night as the astrometry reference frame AR were
selected, and the transformed magnitude data were averaged (for each
star) with weights being the rms values of the fits. This resulting
master photometry reference (MR) file was then used for re-fitting all
the magnitude files to the MR. This final step reduced our fitting rms
to 2\,mmag for the best frames, and in general by a factor of 30\%.

\vskip-1cm
\section{Attainable precision}
\label{sec:photprec}

We derive the photometric precision of the system by looking at the rms
variation of light-curves of stars we believe are constant. These are
difficult to select {\em a priori}, but it is a good assumption that
majority of the stars are not variable above the millimagnitude level,
and thus we can characterize the photometric precision by the median of
the rms of the magnitude variations within a domain of parameters on which
photometric precision depends. The principal parameter is the magnitude
of the sources, but a magnitude vs.~rms plot with wide-field
instruments yields a somewhat broader distribution of points, due to
the dependence of rms on other parameters, such as radial distance from
the field center (due to vignetting).
Both observational rms variations and intrinsic stellar variability
depend on the timescale of our investigation; we distinguish between
nightly, and long-term precision.  The latter is typically inferior to
the former, because it is affected by instrumental drifts, seasonal
variations of the weather, inclusion of non-photometric nights, and
long-term variability of some of the stars.
Photometric precision also depends on the time-resolution of the data.
Binning of course will reduce the rms of the light-curve
to the extent that the error of individual data points is random rather
than systematic. The following discussion assumes the 5-min
time-resolution of HAT observations, unless noted otherwise.  It is
worth noting that while the signal-to-noise ratio (SNR) of a
planetary transit detection is basically unaffected by binning, the
higher precision of binned data points can be useful in confirming the
reality of the transit.

Comparison of precision with tracking and PSF-broadened frames was
established on photometric nights by taking alternating 5min exposures
of the same field with the two methods. Although this paper deals
mainly with HAT-5, we also present the results of the same experiment
performed simultaneously to HAT-5 on a different stellar field with the
essentially identical HAT-6 telescope, located at the same site. This
way we can rule out possible detector-related issues, and because the
gain by PB is greater for HAT-6, it makes a useful comparison. The
difference between HAT-5 and HAT-6 lies in the intrinsic profile
widths: for HAT-5 this width is 1.7\,pix, strongly variable over the
field, while for HAT-6 the PSF is 1.2\,pix wide. (This is probably due
to the differences between the seemingly identical lenses.) The
PSF-broadening widens the profiles to 2.3\,pix on HAT-5, and to
2.0\,pix on HAT-6. Light-curves were constructed from both series
(tracking and PB) of images, and the I-magnitude vs.~rms curves were
compared. Naturally, due to the different PSF-widths with tracking and
PB, the optimal parameters (aperture, sky annulus) for photometry were
different, so we compare the {\em best} photometry we could achieve
with the given technique. Tracking and PB aperture photometry results
were transformed to the same instrumental system using $\sim 1000$
stars, with rms of the transformation smaller than 0.01mag, so that
both sets of data are plotted on the same magnitude scale. Following
this, we performed crude absolute calibration to I-band, using
$\sim$~20 non-saturated Hipparcos \citep{HIP} stars in each field.  The
rms of these transformations is surprisingly high: $\sim 0.2$ mag for
each field, although this does not affect our conclusions when
comparing PB to tracking, which are very precisely in the same
instrumental system.

\begin{figure}[!h]
\begin{center}
\epsscale{0.9}
\plotone{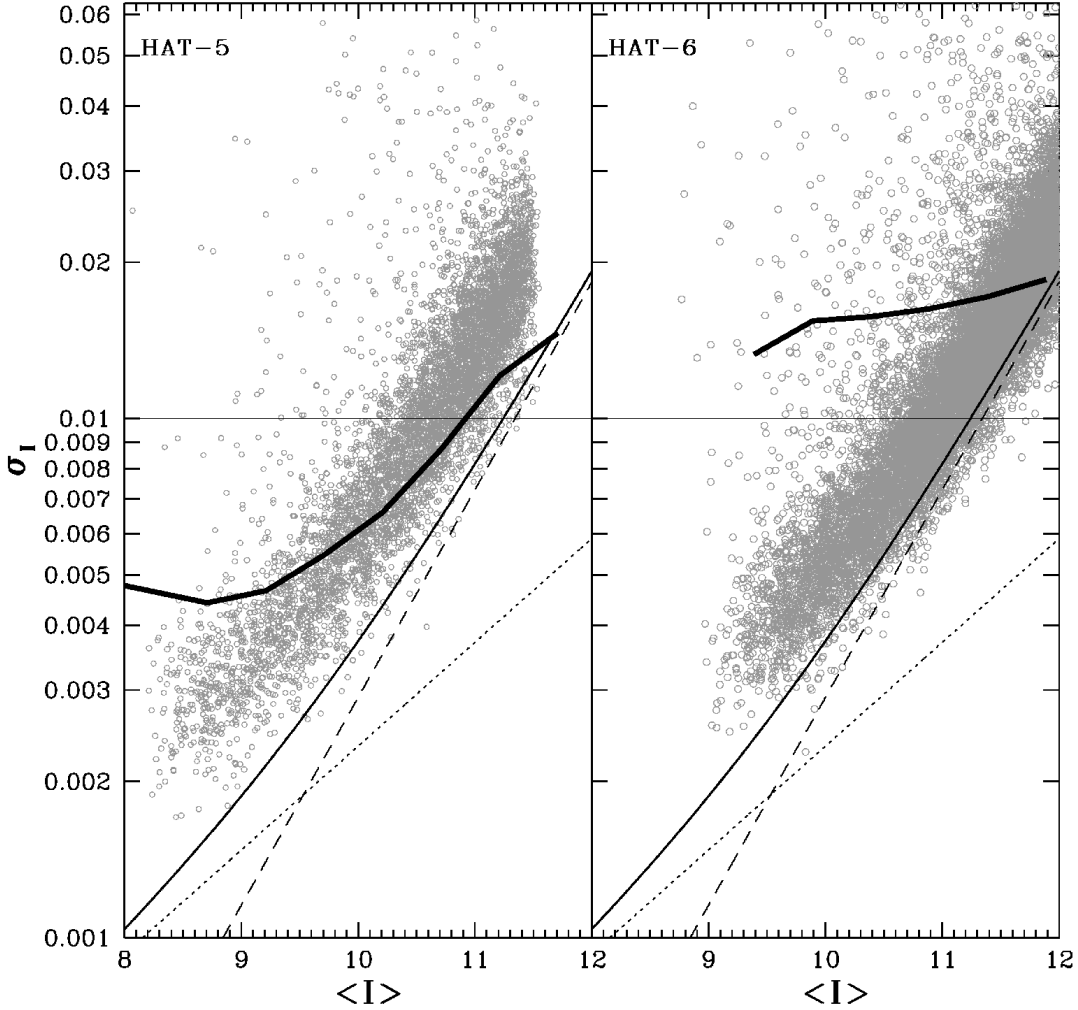}
\caption{Photometric precision attained with HAT-5 (left panel) and
HAT-6 (right), as established from the rms of the light-curves with
5min time-resolution, after 3 sigma rejection of outlier points. The
thick solid line represents the median rms values in 0.25mag bins from
simple tracking frames, while the open circles show the improvement by
PB for each star, reaching 2\,mmag for the brightest stars. The
difference is enhanced at HAT-6, where the intrinsic PSF is very sharp
(1.2\,pix) compared to the PB (2.0). Theoretical noise estimates are
also given for the PB-case (3.0\,pix aperture): the combined noise
(thin solid line) is the sum of sky-noise (dashed line) and
photon-noise (dotted line).
\label{fig:photprec}
}
\end{center}
\end{figure}

\notetoeditor{Fig.~\ref{fig:photprec} needs to be 2column wide,
definitely not shrunk to 1column. The data points are grayscale so as
not to cover the thick lines.}

Fig.~\ref{fig:photprec} compares the precision attainable when
employing straight tracking or PB.  For both HAT-5 and HAT-6, the
precision on the bright end is considerably better with PB (grayscale
open circles) than with tracking (median value only shown as thick
solid lines, to avoid confusion of overlapping individual points).
However the precision using PB is worse for faint stars. The poorer
behavior on the faint end can be explained by the effect of increased
sky-background noise under the widened profile of stars with small
flux. The dramatic improvement in precision on the bright end is
especially significant for HAT-6 (right), which has sharper intrinsic
profiles. (The fact that stars extend to brighter magnitudes at HAT-5 is
due only to the technical detail that the AR used for HAT-5 data
reduction of these test observations was established on frames with
wider intrinsic PSF, thus containing brighter non-saturated coordinates
for fixed-center photometry. Saturated sources, because of the
coordinate uncertainties of their profiles, were rejected from the
coordinate lists in this comparison.)
It may be noted that the above-mentioned rms uncertainty of $\sim 0.2$
mag in the transformation from instrumental to true I-band magnitudes
means that the vertical axes of the left and right panels of
Fig.~\ref{fig:photprec} must be considered uncertain relative to each
other by perhaps as much as 0.1 mag.

The gain in precision when PB is used can be attributed to a number of
causes: The residual inter-pixel variations of imperfect flatfielding
(and to smaller extent, the intra-pixel variations) are somewhat
smoothed out by the broader PSFs.  Because of the broader PSF, and use
of brighter non-saturated stars with higher S/N, the resulting
magnitude transformation between the frames is more accurate. Because
the optimal aperture for PB is wider, the errors in the centroid
position of stars yield smaller errors in photometry.  Finally, the
more homogeneous profiles with time and position within the field act
to maintain the stability of the measured magnitudes over pointing
changes or focus changes; at the same time the magnitude transformation
is simpler than for tracking frames.

Theoretical expectations on the noise sources are also shown on
Fig.~\ref{fig:photprec} (thin solid lines), using the standard formulae
described in e.g.~\citet{Newberry91} that depend on the incoming flux
(in \el), gain of the CCD, number of pixels in the aperture and
background annulus, and the standard deviation of the background (as
measured from reduced frames). It may be seen that the theoretical
curves describe the empirical results of PB fairly well, but with the
undersampled (pure tracking) profiles other effects start to dominate,
and the measured rms values (median showed as thick solid lines)
diverge from the theory: they flatten on the bright end, at values
(0.5\% for HAT-5 and greater than 1\% for HAT-6) that depend on the
width of the intrinsic PSF.

In summary, while precision in the star-noise limited regime is
drastically improved, the precision becomes a steeper function of
magnitude, and degrades rapidly for faint sources in the sky-noise
limited regime. The actual performance depends on the intrinsic and
broadened PSF widths. The PB-technique provides another degree of
freedom when optimizing the photometric characteristics of a system.
The advantages are determined by the goals; if the purpose is to
optimize precision for faint sources (even at the cost of losing
precision for bright stars) or to improve detection of larger amplitude
variability for very faint sources (as was the case for the previous
use of these detectors and lenses in the ROTSE-I project,
\citealt{Akerlof00}), then the sharpest possible profiles are desired. 
PSF-broadening is not a general solution for improving precision of a
system, rather a special technique tested on our instrumentation, which
increases the number of stars having photometry better than 1\%, and
thus the chance of detecting planet transits.  For wide-field planet
searches such as ours, which use semi-professional, front-illuminated
CCDs, if precision for bright stars flattens at an undesired $\sim 1\%$
value, PB might be a workaround.

\subsection{Long-term variation of precision}

\begin{figure}[!h]
\begin{center}
\epsscale{0.7}
\plotone{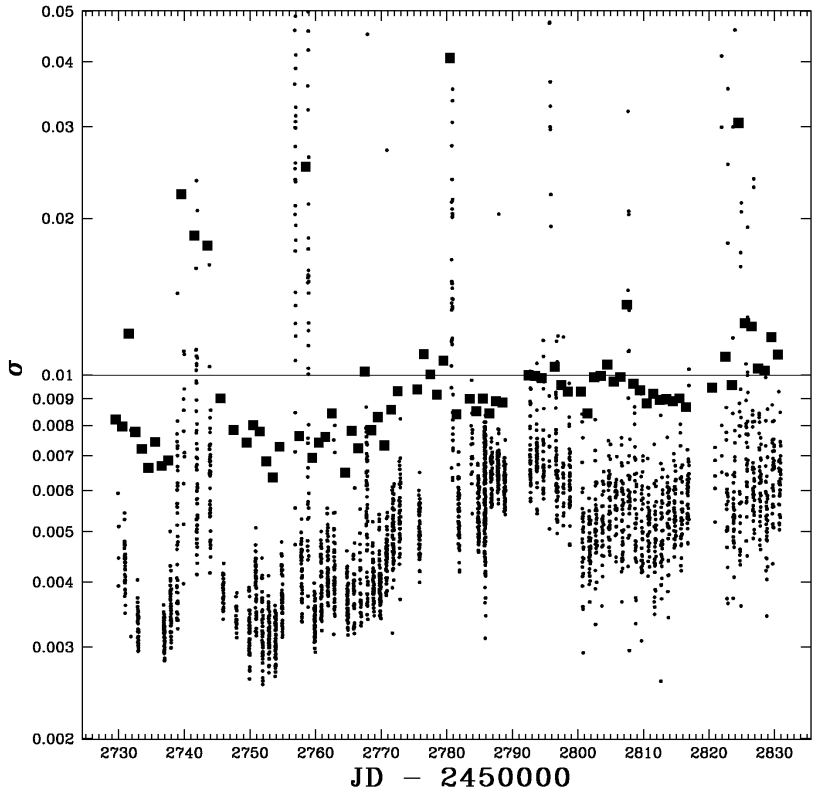}
\caption{
Long term photometric precision of HAT-5 in the Spring 2003 season, as
a function of Julian date. Both the nightly rms variations of
$I=10\pm0.25$ magnitude stars (big filled boxes) and the rms of the
magnitude transformation to the reference for each frame (small dots)
are plotted. Beside the outlier points due to non-photometric nights, a
long-term degradation is also visible.
\label{fig:longprec}
}
\end{center}
\end{figure}

Over the 4 months of operation in the 2003 Spring season, we
experienced a long-term degradation of the nightly precision, as is
shown in Fig.~\ref{fig:longprec}, where we plot the nightly rms
variations of stars of similar brightness ($I=10\pm0.25$, $\sim$600
sources) as a function of JD (big filled boxes). This was later found
to be due to the PSFs becoming increasingly ``over-broadened'', because
of a tight declination-drive sprocket on HAT-5, and possibly other
instrumental misalignments.  The outlier points are due to
non-photometric nights.  The rms of the magnitude transformation to the
photometric reference is also plotted for each frame (small dots,
$\sim$80 per night), and its correlation with the actual nightly rms of
stars can be used as a direct measure of the photometric conditions. We
stress that the data reduction methodology discussed earlier ensures
that the stability of magnitudes over several months is comparable to
the rms of magnitude-fits from individual frames to our photometric
reference.

Fourier analysis of the brightest 2000 light-curves (with rms typically
less than 1\%) revealed other effects, which potentially harm planet
and low-amplitude variability detection capabilities. These are
miscellaneous low-amplitude ($\sim1\%$, always smaller than 0.1mag)
systematic variations exhibited by most of the stars, the exact subset
of stars depending on the effect itself. The fact that the variations
are not intrinsic to the stars is seen from either the frequency: daily
variations close to 1, 1/2 and 1/3 $d^{-1}$, or by peaks in the
frequency-distribution histogram: 38min and 80min modulations. Although
we have indications about the origin of these systematics, they haven't
been completely understood and elimination poses a serious task to be
solved. Some are definitely related to the minor pointing imperfections
of the telescope, such as periodic tracking errors (with 38min
periodicity), and are expected to diminish with the introduction of the
position auto-correction technique described in \S \ref{sec:hw} (in use
since the Fall of 2003). It is especially intriguing to note that some
stars show a daily variation, while others have stable light-curves,
and yet for the stars showing this variation no clear dependence on
other parameters (color, position) have been found. The various
modulations seem to be related, since pre-whitening with the 1\,day
periodicity usually causes the 80min component almost to disappear. We
note that various systematics are also observable in other large scale
surveys, such as MACHO \citep{Alcock00} and OGLE \citep{oglesys}.

\section{Early results}
\label{sec:res}

\subsection*{Search for planetary transits}

To search for possible planetary transits, a BLS period search 
\citep{GK02}  was performed on the moderately rich Hercules field, 
which had $\sim3000$ observations, more than the twice the number of
observations of the sparse Sextans field. The search was limited to the
[0.02,0.98]$d^{-1}$ frequency range with 20000 frequency steps, 200
phase bins, fractional transit lengths of $q_{min}=0.005$ and
$q_{max}=0.05$. The BLS spectra generally displayed an increasing
background power toward lower frequencies, most probably because of
slight long-term systematic trends of the light-curves. Therefore, we
fitted 4th order polynomials to the spectra by using 5 sigma clipping
so as to minimize the effect of large peaks. Then, these polynomials
were subtracted from the spectra and the residuals were examined for
outstanding peaks. The BLS algorithm, similarly to other matching
methods, creates subharmonics, and the 1day systematic variation of
certain stars resulted in false frequency peaks at k/n [$d^{-1}$],
where k and n are small integers. The distribution function of the peak
BLS frequencies clearly showed such peaks, and stars exhibiting these
periodicities were excluded.

\begin{figure}[!h]
\begin{center}
\plotone{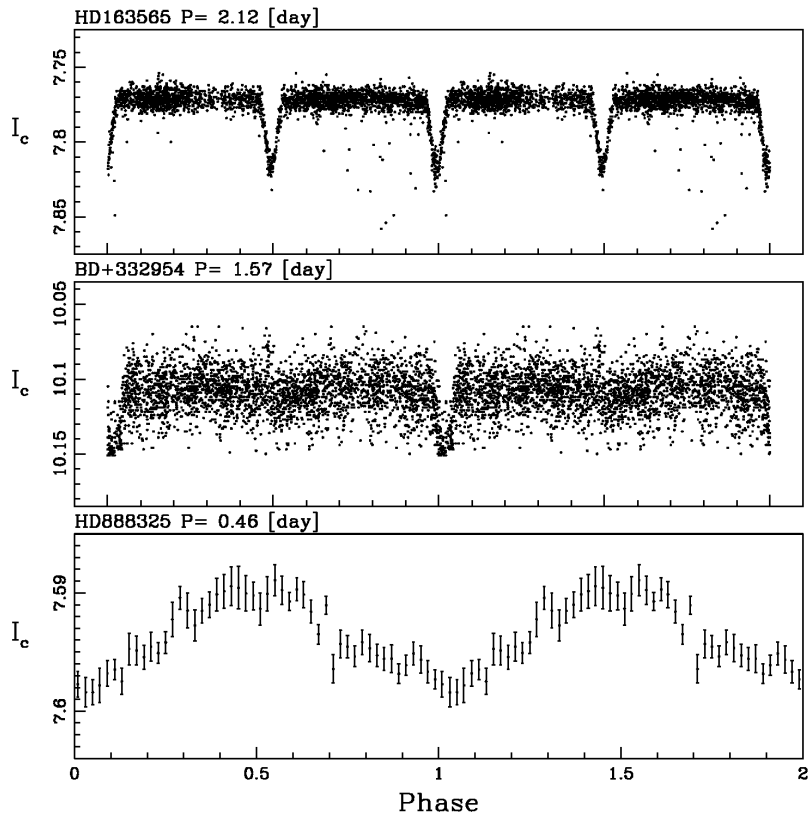}
\caption{
Phased light-curves of three variables of special interest. 
Upper panel: Star \mbox{HD 163565} shows V-shaped photometric dips of 4\%
depth, characteristic of grazing eclipses, with full period (two nearly
identical dips) of 2.11 days.
Middle panel: Star \mbox{BD+33 2954} with F8 spectrum shows a periodic
dip with amplitude of only 0.032 mag, period 1.57 days, and duration
2.0 hrs, perhaps caused by an M-dwarf; there may be a blended
tertiary in the system.
Lower panel: \mbox{HD 88832} is an A5 star with only 8\,mmag
peak-to-peak amplitude of variation.  The light-curve for this star has
been binned in phase with $\sim25$ data points occurring in each bin.
\label{fig:lc}
}
\end{center}
\end{figure}

Finally, a sample of 133 stars was selected from the remaining sources,
having BLS spectral peaks with signal-to-noise ratio (SNR) greater than
6.0. Visual inspection of the folded light-curves further narrowed the
sample to 23 variables, with most detections being marginal. The bottom
line is that no definite planetary transit detections have been
found in this early dataset. The two best cases (Figure \ref{fig:lc}
upper and middle panel) belong to a V-shaped grazing eclipse of 4\%
depth observed for the A0 star \mbox{HD 163565}, and a shallow (3.5\%)
eclipse of the F8 star \mbox{BD+33 2954}, which is too deep to be
planetary transit; the system also shows a low-amplitude sinusoidal
variation in its high-luminosity state. The remaining sources were well
below the significance of the the above two stars.

We also performed tests on the detectability of periodic transit events
on the Hercules field. The present figures for the detection rates are
preliminary, as the technique for recovering signals is under
development. A transit of P=$5.12345^d$, dip=$-0.015$ mag, length =
$0.03P$ was added to the brightest $\sim2000$ observed light-curves
($I\lesssim10.5$). This way the recovery was tested on the real
observations, not even excluding sources that show variability (the
fraction of the latter is approx.~10\% or less, see later). The BLS test
was run on the time-series with search parameters described in the
beginning of this section. The selection criterion for detection were
i) SNR of the BLS spectrum greater than 6.0, and ii) period
corresponding to the peak in the range of [5.102,5.154] days. The
detection rate is rather high (74\%) on the bright end, especially
after excluding variable stars from the sample (82\%).  However, it
drops to 29\% for the faintest stars at $I\approx10.5$. 

\subsection*{General Stellar Variability}

The few-mmag precision of HAT for bright stars allows for investigating
general stellar variability at levels not normally attainable from the
ground with small, very wide-field telescopes. HAT's capability to
measure thousands of stars in a single field with better than 1\%
precision should enable it to contribute significantly to general
stellar variability studies.

For strictly periodic variations, phase-binning of observations over
many periods allows us to achieve relative precision in a light-curve
far better than that of individual observations; the ultimate limit is
placed by systematic effects that cannot be corrected for. As an
example of phase-binning, Figure \ref{fig:lc} (lower panel) illustrates
a very low-amplitude (8\,mmag peak-to-peak) quasi-sinusoidal
photometric variation of \mbox{HD 88832}, an I=7.6 (V=8.1) A5 star in
the Sextans field. Analysis of Variance \citep[AoV:][]{Sch96} study
revealed a predominant photometric period of 0.462 days. The data (1226
data points over 65 days) were phased to this period and divided into
50 bins, so that approximately 24 data points occur in each. The
average standard deviation of the mean of these data points is
1.1\,mmag, illustrating that the magnitude calibration is stable at
that level over a period of at least two months.

We note that for detection of a planetary transit, the situation is
much less favorable, since a typical transit would occupy only about 3
to 4 percent of the total phase diagram (i.e.~one to two of the phase
bins in Figure \ref{fig:lc} lower panel).

In order to establish a preliminary variable star inventory,
light-curves were searched for using Discrete Fourier Transformation
(DFT) with 5-sigma clipping of potentially bad data-points, and
imposing a minimal SNR of 6.0. Variables with close integer frequencies
were rejected ($f_0 < 0.02 d^{-1}$ and $n-0.018 < f_0 < n+0.018 d^{-1}$
where n is integer). Finally, folded light-curves were subject to
visual inspection and rejection. This preliminary analysis definitely
lost low-amplitude variables hidden in the systematic trends, or those
with multiple periods, and it is also incomplete at long periods that
are comparable to the time-span of observations. Nevertheless, most
variables with greater than 0.01mag amplitude were recovered, and the
total number turned out to be $\sim220$ (11\%) in the Hercules field,
and $\sim 110$ (5.5\%) in the Sextans field, respectively, out of the
2000 brightest stars tested. This difference is partly because of the
smaller number of observations for the Sextans field (1300 vs.~3300),
but might also be credited to different stellar populations tested. 

\section{Summary and future prospects}
\label{sec:sum}

Photometric precision substantially better than 1\% is important for
detection of giant planets transiting sun-like stars, and also can
contribute significantly to general stellar variability research. We
have discussed the difficulties in attaining this level of precision
when using an instrument such as HAT, i.e.~using a lens with a fast
focal ratio and wide field, attached to a front-illuminated CCD. Most
of these difficulties can be dealt with by careful design of the
instrumentation, specialized observing techniques, and custom-written
software. The problem of under-sampling of stars needs special
treatment. We described our PSF-broadening technique which greatly
improved the precision, reaching 2\,mmag rms for stars at
$I\approx8$. We attribute the gain in precision to the smoothing of
residual inter- and intra-pixel variations caused by imperfect
short-scale flatfielding, the use of brighter stars for matching the
magnitude system of frames, and the better behavior of existing
photometry software on broader PSFs.

HAT-5 has been operational since early Spring, 2003, and the initial
few months were devoted for tests. Based on the observations spanning
four months, 91 nights in total, we determined the long-term precision
and investigated low-amplitude systematic variations. Our search for
periodic transits, performed on both of our observed fields, has not
yielded any definite detection of planetary transits. A few shallow
transits and grazing eclipsing binaries were revealed, along with
several hundred variable stars.

In order to increase the chance of detecting transiting planets, a
network of HAT telescopes (HATnet) has been developed following the
installation of our prototype HAT-5. Two additional systems,
essentially identical to HAT-5 (HAT-6 and HAT-7) were installed at
FLWO, and have been operational since September, 2003. Two more, also
identical, HAT telescopes have now been installed at SAO's Submillimeter
Array site on Mauna Kea, Hawaii, and have been observing since
November, 2003.  A great advantage of the HATnet, presently consisting
of five stations, is that the instruments have nominally identical
telescope mounts, software environment, lenses and CCDs. The longitude
separation of 3 hours between FLWO and Hawaii, and uncorrelated weather
patterns significantly extend our time coverage of fields. We hope that
continuous operation of HATnet and further improvements to be
introduced in our data reduction and analysis software system will
further enhance our ability of transit detection and to study
low-amplitude stellar variability in general.

\acknowledgements
We are greatly indebted to Bohdan Paczy\'nski for initiating
development of HAT. We are grateful to Irwin Shapiro, Eugene Avrett and
the other associate directors of the Center for Astrophysics for
providing internal funds that allowed us to establish the five-element
network. Additional support was provided through NASA Grant NAG5-10854.
We are greatly indebted to Carl Akerlof and the University of Michigan
for loaning us the CCDs and lenses of the decommissioned ROTSE-I
project. Installation and operation of the HAT telescopes at FLWO were
strongly supported by Emilio Falco, Robert Kirshner, Daniel Fabricant,
and telescope operators Perry Berlind and Michael Calkins. We are
grateful to James Moran and Anthony Schinckel for supporting
installation at SAO's SMA site at Hawaii. We wish to thank the HAT
engineers I.~Papp, J.~L\'az\'ar and P.~S\'ari from Hungary for their
invaluable contribution to the development and installations.
Development of HAT also profited from helpful conversations with Andrew
Szentgyorgyi. G.B and G.K wish to acknowledge contribution of grant
OTKA-38437.



\begin{thebibliography}{}

\bibitem[Akerlof et al.(2000)]{Akerlof00} Akerlof, C.~et al.\ 
2000, \apjl, 532, L25 

\bibitem[Alcock et al.(2000)]{Alcock00} Alcock, C.~et al.\ 2000, 
\apj, 542, 257 


\bibitem[Bakos et al.(2002)]{GB02} 
Bakos, G.~{\' A}., L{\' a}z{\' a}r, J., Papp, I., S{\' a}ri, 
P., \& Green, E.~M.\ 2002, \pasp, 114, 974 

\bibitem[Bessell(1990)]{Bessel90} Bessell, M.~S.\ 1990, \pasp, 
102, 1181 

\bibitem[Brown et al.(2001)]{TB01} Brown, T.~M., 
Charbonneau, D., Gilliland, R.~L., Noyes, R.~W., \& Burrows, A.\ 2001, 
\apj, 552, 699 

\bibitem[Brown(2003)]{TB03} Brown, T.~M.\ 2003, \apjl, 593, 
L125 

\bibitem[Buffington, Booth, \& Hudson(1991)]{Buf91} 
Buffington, A., Booth, C.~H., \& Hudson, H.~S.\ 1991, \pasp, 103, 685.

\bibitem[Charbonneau et al.(2000)]{DC00} 
Charbonneau, D., Brown, T.~M., Latham, D.~W., \& Mayor, M.\ 2000, \apjl, 
529, L45 

\bibitem[Charbonneau et al.(2002)]{DC02} 
Charbonneau, D., Brown, T.~M., Noyes, R.~W., \& Gilliland, R.~L.\ 2002,
\apj, 568, 377


\bibitem[Chromey \& Hasselbacher(1996)]{Chromey96} Chromey, 
F.~R.~\& Hasselbacher, D.~A.\ 1996, \pasp, 108, 944. 

\bibitem[Groth(1986)]{Groth86} Groth, E.~J.\ 1986, \aj, 91, 1244 

\bibitem[Henry et al.(2000)]{Henry00} Henry, G.~W., Marcy, G.~W., 
Butler, R.~P., \& Vogt, S.~S.\ 2000, \apjl, 529, L41 

\bibitem[Horne(2003)]{KH03} Horne, K.\ 2003, ASP 
Conf.~Ser.~294: Scientific Frontiers in Research on Extrasolar Planets,
eds.~D.~Deming and S.~Seager, San Francisco,
The Astronomical Society of the Pacific, 361 
%

\bibitem[Konacki et al.(2003)]{Konacki03} 
Konacki, M., Torres, G., Jha, S., \& Sasselov, D.~D.\ 2003, \nat, 421, 507


\bibitem[Kov{\' a}cs, Zucker, \& Mazeh(2002)]{GK02}
Kov{\' a}cs, G., Zucker, S., \& Mazeh, T.\ 2002, \aap, 391, 369

\bibitem[Kruszewski \& Semeniuk(2003)]{oglesys} Kruszewski, 
A.~\& Semeniuk, I.\ 2003, Acta Astronomica, 53, 241 

\bibitem[Landolt(1992)]{Landolt92} Landolt, A.~U.\ 1992, \aj, 
104, 340

\bibitem[Lasker et al.(1990)]{GSC} Lasker, B.~M., Sturch, 
C.~R., McLean, B.~J., Russell, J.~L., Jenkner, H., \& Shara, M.~M.\ 1990, 
\aj, 99, 2019. 

\bibitem[Newberry(1991)]{Newberry91} Newberry, M.~V.\ 1991, \pasp, 
103, 122 

\bibitem[Perryman et al.(1997)]{HIP} Perryman, M.~A.~C.~et 
al.\ 1997, The Hipparcos and Tycho Catalogues, ESA SP-1200, Vol.~1-17

\bibitem[Pojma{\' n}ski(1997)]{GP97} Pojma{\' n}ski, G.\ 1997, Acta 
Astronomica, 47, 467.

\bibitem[Queloz et al.(2000)]{Queloz00} Queloz, D., Eggenberger, 
A., Mayor, M., Perrier, C., Beuzit, J.~L., Naef, D., Sivan, J.~P., \& Udry, 
S.\ 2000, \aap, 359, L13 

\bibitem[Schwarzenberg-Czerny(1996)]{Sch96} 
Schwarzenberg-Czerny, A.\ 1996, \apjl, 460, L107 

\bibitem[Shewchuk (1996)]{Del96} Shewchuk, R.~J.~1996, in
Applied Computational Geometry:  Towards Geometric Engineering, 
eds.~M.~C.~Lin and D.~Manocha, Berlin, Springer-Verlag, 
vol.~1148, 203, 
http://www-2.cs.cmu.edu/\~{ }quake/triangle.research.html and
references therein

\bibitem[Torres et al.(2003)]{Torres03}
Torres, G., Konacki,D. M., Sasselov, D.~D. \& Jha, S.\ 2003, astro-ph
0310114, submitted to \apjl

\bibitem[Udalski et al.(2002)]{Udalski02} Udalski, A., Zebrun, 
K., Szymanski, M., Kubiak, M., Soszynski, I., Szewczyk, O., Wyrzykowski, 
L., \& Pietrzynski, G.\ 2002, Acta Astronomica, 52, 115 

\bibitem[Vidal-Madjar et al.(2003)]{VM03} Vidal-Madjar, A., 
Lecavelier des Etangs, A., D{\' e}sert, J.-M., Ballester, G.~E., Ferlet, 
R., H{\' e}brard, G., \& Mayor, M.\ 2003, \nat, 422, 143 

\bibitem[Wren et al.(2001)]{ROTSEI} Wren, J.~et al.\ 2001, 
\apjl, 557, L97 

\end{thebibliography}
\end{document}